\documentclass[11 pt,oneside]{article}
\usepackage[top=1in, bottom=1.2in, left=1.1in, right=1.1in]{geometry}
\usepackage{setspace, color, natbib, bm, amsmath, amsthm, amssymb,amsbsy,amsfonts,mathrsfs, float, appendix, authblk, graphicx, caption, enumitem, booktabs, titlesec,soul,xcolor, caption, subcaption, threeparttable, lscape, adjustbox, multirow, gensymb}

\usepackage[breaklinks=true]{hyperref}

\csname @openrightfalse\endcsname

\setlist{nolistsep}
\makeatletter
\newcommand*{\rom}
[1]{\expandafter\@slowromancap\romannumeral #1@}
\makeatother
\hypersetup{
     colorlinks   = true,
     citecolor    =  blue
}

\hypersetup{
    colorlinks=true,
    linkcolor=blue,
    filecolor=magenta,      
    urlcolor=cyan,
    pdftitle={Overleaf Example},
    pdfpagemode=FullScreen,
    }

\begin{document}

\title{\textbf{Trends in Temperature Data: Micro-foundations of Their Nature}}
\author[1]{Maria Dolores Gadea-Rivas}
\author[2]{Jesús Gonzalo}
\author[3]{Andrey Ramos}
\affil[1]{Department of Applied Economics, University of Zaragoza, lgadea@unizar.es}
\affil[2]{Department of Economics, Universidad Carlos III de Madrid, jesus.gonzalo@uc3m.es}
\affil[3]{Department of Economics, Universidad Carlos III de Madrid, anramosr@eco.uc3m.es}

\clearpage\maketitle

\begin{abstract}

Determining whether Global Average Temperature (GAT) is an integrated process of order 1, I(1), or is a stationary process around a trend function is crucial for detection, attribution, impact and forecasting studies of climate change. In this paper, we investigate the nature of trends in GAT building on the analysis of individual temperature grids. Our 'micro-founded' evidence suggests that GAT is stationary around a non-linear deterministic trend in the form of a linear function with a one-period structural break. This break can be attributed to a combination of individual grid breaks and the standard aggregation method under acceleration in global warming. We illustrate our findings using simulations. 



\end{abstract}

\section{Introduction}

Global Average Temperature (GAT) observed since the late 1800s exhibits an upward trend widely interpreted as evidence of global warming \citep{Mann, gadeagonzalo2020, ipcc}. The specific nature of the trend is an open question in the empirical literature with a non-trivial answer. Neither theoretical climate nor economic models with climatic variables help to identify a particular trend specification. From the climate quantitative perspective, two dominant strands of literature with opposing views debate. On one side, authors as \cite{Woodward}, \cite{kaufman2010}, \cite{dergiades2016}, \cite{bruns2020}, and \cite{phillips2020} approximate the GAT series by an integrated process of order 1, I(1), and advocate for the use of modelling approaches accounting for the presence of stochastic trends. On the opposing side, \cite{seidel}, \cite{gaygarcia2009}, \cite{estrada2013}, \cite{estradaperron2017}, \cite{mudelsee}, or \cite{mcKitrick} defend that GAT can be best represented as trend-stationary process with a one-period structural break in the trend function. The assumption of deterministic trends is also common in the climate analysis as inferred from the statistical procedures implemented by the Intergovernmental Panel on Climate Change (IPCC) on its Sixth Assessment \citep{ipcc} and previous reports.

Determining whether GAT is an I(1) process or a stationary process around a non-linear trend is a key issue for detection, attribution, and impact studies of climate change (see \citealt{mcKitrick}). If GAT is assumed to be I(1), attribution of global warming involves demonstrating that the stochastic trends in temperature are inherited from its association with the anthropogenic forcing from CO2 and other greenhouse gases.\footnote{\cite{Pretis_Hendry} and \cite{Bennedsen_gcb} study the statistical properties of radiative forcing from different components (including CO2) and find that the series follows an I(1) or even an I(2) process.} However, the I(1) assumption also implies that exogenous temporary shocks like solar flares or volcanic eruptions generate long-lasting effects on temperature, which does not seem to be the case in the observed record.\footnote{Notice that a shock with transitory effects on GAT can have permanent effects on other variables.} 

If trends are assumed to be deterministic, detection of global warming is possible through traditional trend-tests (see \cite{gadeagonzalo2020} for a robust trend-detection analysis). But this assumption poses challenges for attribution and impact studies due to the problem of 'unbalanced' relations. Concretely, in attribution studies,  the mismatch in the order of integration between GAT and anthropogenic forcing hinders the estimation of the climate sensitivity parameter (for a review, see \cite{Rohling}) using regression analysis. This is also the case for the impact studies that rely on dynamic growth equations linking the growth rate of per-capita output with temperatures (see \cite{dell} and references therein). Beyond this literature, understanding the nature of the trends is helpful for producing more accurate long-term forecasts as stressed by \cite{kaufman2010}.

The existing evidence regarding the nature of trends in GAT relies on the analysis of aggregated series. This paper contributes to the debate by offering a 'micro-founded' explanation for the observed trends based on the study of the trend dynamics of individual units used in computing those averages. On aggregate, our findings are consistent with the hypothesis of \cite{seidel} and \cite{gaygarcia2009}, suggesting that temperature averages are trend-stationary with a one-time permanent shock breaking the trend function. Unit root tests implemented on individual grids provide evidence in the same direction. Our empirical analysis points towards the standard aggregation process as being important in explaining the origin of trend-breaks in aggregated series when combined with an accelerated global warming. Depending on the strength of the break's signal, these situations may bias standard unit root tests towards the non-rejection zone. We illustrate our hypothesis through a set of Monte-Carlo simulations assuming linear and broken-trend individual processes and emulating the standard aggregation methods used to compute the GAT.

The rest of the paper is organized as follows. Section 2 provides the empirical evidence regarding trends in aggregated and individual temperature series. Section 3 presents our simulation exercises. Finally, Section 4 concludes.

\section{Empirical Evidence}

\subsection{Data}

\begin{figure}[H]
     \centering
     \begin{subfigure}{0.49\textwidth}
         \centering
         \includegraphics[width=\textwidth]{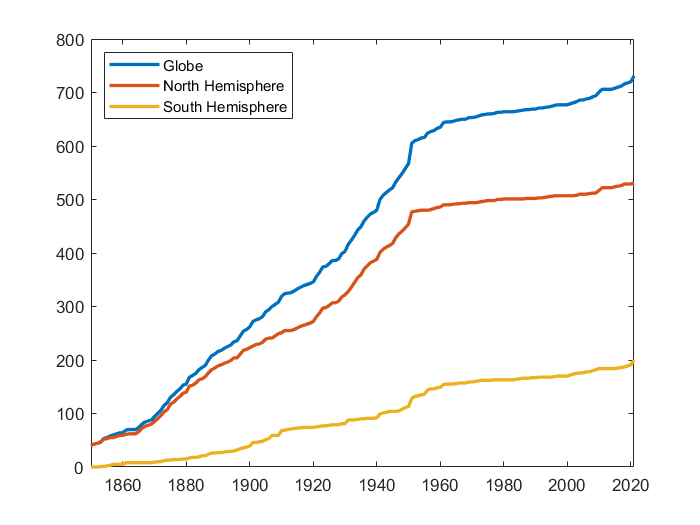}
         \caption{Number of grids observed continuously}
         \label{num_observed}
     \end{subfigure}
     \hfill
     \begin{subfigure}{0.49\textwidth}
         \centering
         \includegraphics[width=\textwidth]{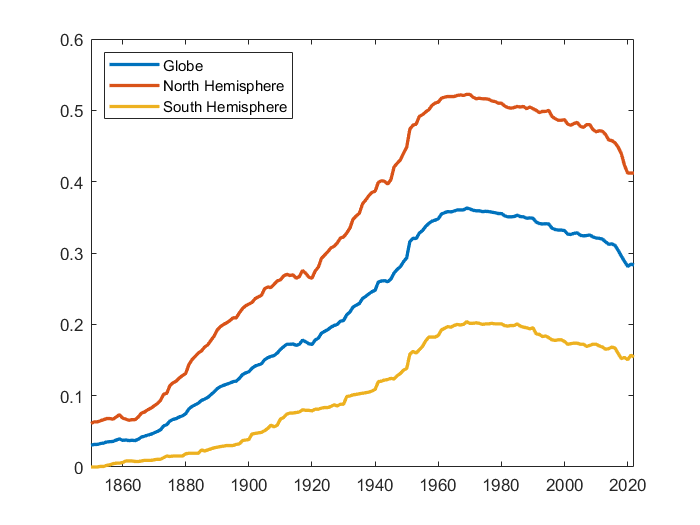}
         \caption{Proportion of non-missing grids}
         \label{prop_observed}
     \end{subfigure}
   \caption{Observation process for the gridded temperatures} 
\label{fig_observation}
\end{figure}

Temperature data is obtained from the latest version of the HadCRUT5 dataset, jointly developed by the Climatic Research Unit (CRU) at the University of East Anglia and the Hadley Centre at the UK Met Office.\footnote{The dataset is accessible at the following URL: \url{https://crudata.uea.ac.uk/cru/data/temperature/}.} For land regions across the globe, this dataset compiles monthly-mean temperatures from a network of 10,000 weather stations, spanning the period from 1850 to 2022. Out of these stations, almost 8,000 are used to construct gridded observations at a resolution of $5 \degree \times 5 \degree$. Data coverage is denser over the more populated regions, particularly, the United States, southern Canada, Europe, and Japan. In contrast, coverage is more limited over the interior of South America and Africa, as well as the Antarctica. To align with standard aggregation method followed by CRU, in our empirical analysis we use directly the series of gridded temperature anomalies from the period 1961-1990.\footnote{Similar results are obtained if we use raw-stations data.} A feature of the dataset is that the number of grids with non-missing data is relatively low during the early part of the record and gradually increases over time. Panel (a) of Figure \ref{fig_observation} presents the number of grids that are continuously observed from the given year onwards, while Panel (b) plots the proportion of non-missing grids (out of 2,592) each year. For a more detailed explanation on the construction of grids, see \cite{morice}.

\subsection{Computing Average Temperature}

Two alternative methods to compute average temperatures are considered. Method A aggregates the full set of grids with non-missing data available each year. Denoting the resulting average as $\bar{T}_{t}^{A}$, for each $t$:
\begin{equation}
    \bar{T}_{t}^{A} = \frac{1}{N_{t}}\sum_{i = 1}^{N}I_{it} \times T_{it}, \hspace{0.3cm} t=1,...,T,
    \label{eq_methodA}
\end{equation}
where $T_{it}$ is the temperature in grid $i$ at year $t$, $I_{it}$ is an indicator for $T_{it}$ non-missing, and $N_{t} = \sum_{i=1}^{N} I_{it}$. Due to the non-uniform observation process for grids, this approach uses data from more grids in recent years compared to the earlier part of the record. i.e., $N_{t}$ grows with $t$. Method A closely resembles the standard aggregation procedure adopted by CRU.

Method B, on the other hand, selects the set of grids with non-missing data throughout the entire sample period, ensuring a stable number of grids on the computation of the average. Denoting the resulting average as $\bar{T}_{t}^{B}$, for each $t$:
\begin{equation}
    \bar{T}_{t}^{B} = \frac{1}{|S|}\sum_{i \in S} T_{it}, \hspace{0.3cm} t=1,...,T,
\end{equation}
where $S = \{i: I_{it}=1, \hspace{0.3 cm} \forall t \}$ and $|S|$ is the cardinality of $S$. This is the approach adopted by \cite{gadeagonzalo2020} to estimate not only the mean but any distributional characteristic of interest.

\begin{figure}[H]
     \centering
     \begin{subfigure}{0.49\textwidth}
         \centering
         \includegraphics[width=\textwidth]{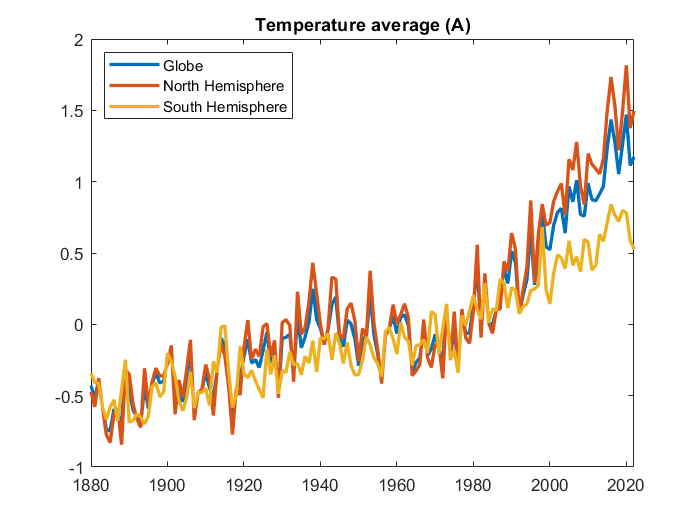}
         \caption{Method A}
         \label{fig_average_A}
     \end{subfigure}
     \hfill
     \begin{subfigure}[b]{0.49\textwidth}
         \centering
         \includegraphics[width=\textwidth]{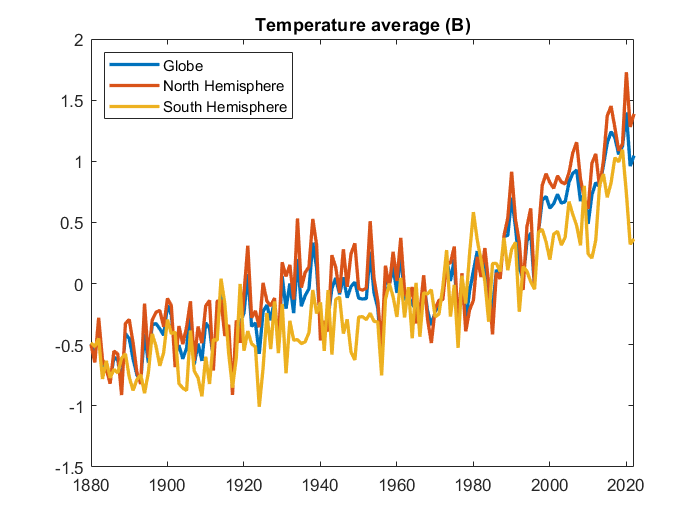}
         \caption{Method B}
         \label{fig_average_B}
     \end{subfigure}
\caption{Average Temperature Series (1880-2022)}
\label{fig_average_1880_2022}
\end{figure}

Separate averages are calculated for the Northern Hemisphere (NH) and Southern Hemisphere (SH), using data from 1880 to 2022.\footnote{Only 155 grids are observed over the full sample period, with the majority of them located in the NH.} Global temperature is obtained as a weighted average of both hemispheres, with the weights accounting for the difference in land areas. The estimated series, presented in Figure \ref{fig_average_1880_2022}, exhibit increasing trends that are indicative of global warming. However, two distinctions emerge depending on the estimation approach. First, the series in Panel (b) exhibits higher variability explained by the smaller number of series used each year to compute averages under method B. Second, in Panel (a), the average in the SH follows a distinct trend compared to the NH and the globe, specially in the second part of the record; this is not observed in the series of Panel (b). 

\subsection{Trends in Average Temperatures}

The nature of the trends in aggregated temperature series is analyzed using time series methods. We initially implement standard Augmented Dickey Fuller (ADF) tests including an intercept and a linear trend and selecting the number of lags based on the Bayesian Information Criterion (BIC). The test-statistics reported in Table \ref{tab_ur_aggregated} indicate that the null of unit root cannot be rejected for the globe and NH averages estimated under method A.\footnote{This is the main evidence for \cite{kaufman2010} and related literature to approximate the trend process through stochastic trends.} In contrast, the null of unit root is rejected in all aggregated series computed following method B.\footnote{Standard ADF-test implemented on other distributional characteristics (quantiles, skewness, kurtosis, etc.) do not show evidence of unit roots neither.}

A potential shortcoming of the ADF test is its sensitivity to the specification of the deterministic component under the alternative. After \cite{Perron}, it is well known that in the presence of a shift in the trend function, the first order autorregresive coefficient is highly biased towards unity and the unit root null is hardly rejected. To address this situation, we complement our analysis with the \cite{KimPerron} (KP) test allowing for a break in the trend function at an unknown period. As observed in Table \ref{tab_ur_aggregated}, for the globe and NH temperatures computed with method A the null of unit root is now rejected. The break in the trend is detected to occur around 1964. For the other series, the test also detects a break but the conclusion of no-unit roots remain the same as before.\footnote{Alternatively, we considered to model the deterministic component using polynomial trends of order 2 and 3. In both cases, the variation of the series around the trend function rejects the unit root hypothesis.}

\begin{table}[h]
\caption{Test-statistics of the unit root tests in average temperatures (1880-2022)}
\centering
\begin{adjustbox}{width=0.6\textwidth}
\begin{threeparttable}
\begin{tabular}{lcc}
\hline \hline
\textbf{Temperature series} & \textbf{ADF test} & \textbf{KP test} \\ \hline
\multicolumn{3}{c}{\textit{Method A}} \\ \hline
Globe                 & -1.619     & -5.514         \\
Northern Hemisphere   & -0.863     & -4.323         \\
Southern Hemisphere   & -6.972     & -9.189   \\ \hline
\multicolumn{3}{c}{\textit{Method B}} \\ \hline
Globe                 & -4.336     & -7.569         \\
Northern Hemisphere   & -4.413     & -7.680         \\
Southern Hemisphere   & -8.5615    & -9.739         \\ \hline
\end{tabular}
\begin{tablenotes}
\item \tiny \textit{Notes:} The table contains the test-statistics of the unit root tests implemented on each average temperature series. ADF-test equation includes an intercept and a linear trend. Number of lags selected based on the BIC. The KP tests allows for one break in the intercept and the slope. Critical values at the 5\% of significance are -3.444 for the ADF-test and -3.760 for the KP test.
\end{tablenotes} 
\end{threeparttable}
\end{adjustbox}
\label{tab_ur_aggregated}
\end{table}

In summary, the aggregated evidence aligns with the hypothesis of \cite{seidel} and \cite{gaygarcia2009}, who argue that temperature process is trend stationary with a shift in the trend function. Unit roots non-rejected by the standard ADF test for method A averages are the consequence of a misspecified model for the deterministic component under the alternative that ignores potential structural breaks. Even though the series estimated under Method B also present a trend-break, the signal is not strong enough to drive the ADF test to the non-rejection zone. All this evidence implies that the aggregation method may have an impact on the conclusion of the tests. 

\subsection{Trends in Individual Temperature Grids}

In this section, we model the trend dynamics of the individual grids used to compute temperature averages and analyze the consequences of the aggregation method for the emergence of structural breaks. 

Individual tests strongly reject the presence of unit roots. Results reported in Table \ref{tab_ur_individual} indicate that the proportion of rejections in the ADF test is high across the three sample periods considered. For instance, in a panel of 354 grids continuously observed during the period 1920-2022, the null is rejected in 88.17$\%$ of the cases. This proportion increases further when implementing the KP test.\footnote{Panel unit root tests point towards the same conclusion.} 

\begin{table}[h]
\caption{Proportion of rejections in the unit root tests on individual grids}
\centering
\begin{adjustbox}{width=0.5\textwidth}
\begin{threeparttable}
\begin{tabular}{lcc}
\hline \hline
\textbf{Sample} & \textbf{ADF test} & \textbf{KP test} \\ \hline
1880-2022   & 92.90\%     & 96.82\%         \\
1920-2022   & 88.17\%     & 97.98\%         \\
1960-2022   & 90.87\%     & 93.86\%   \\ \hline
\end{tabular}
\begin{tablenotes}
\item \tiny \textit{Notes:} The table reports the proportion of times that the null of unit root is rejected. The number of grids are 155, 346, and 365 for the periods 1880-2022, 1920-2022, and 1960-2022, respectively. ADF-test equation includes an intercept and a linear trend. Number of lags selected based on the BIC. The KP tests allows for one break in the intercept and slope. Tests at 5\% of significance.
\end{tablenotes} 
\end{threeparttable}
\end{adjustbox}
\label{tab_ur_individual}
\end{table}

\begin{figure}[H]
     \centering
     \begin{subfigure}{0.49\textwidth}
         \centering
         \includegraphics[width=\textwidth]{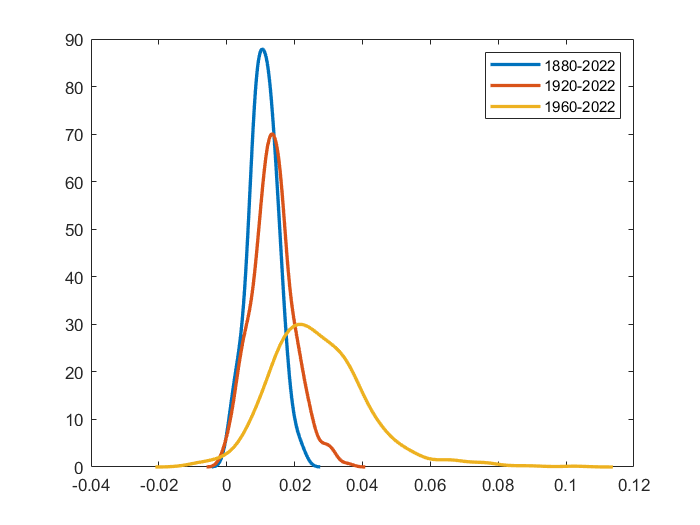}
         \caption{Density of the slopes in the linear trend model}
         \label{fig_density_linear}
     \end{subfigure}
     \hfill
     \begin{subfigure}[b]{0.49\textwidth}
         \centering
         \includegraphics[width=\textwidth]{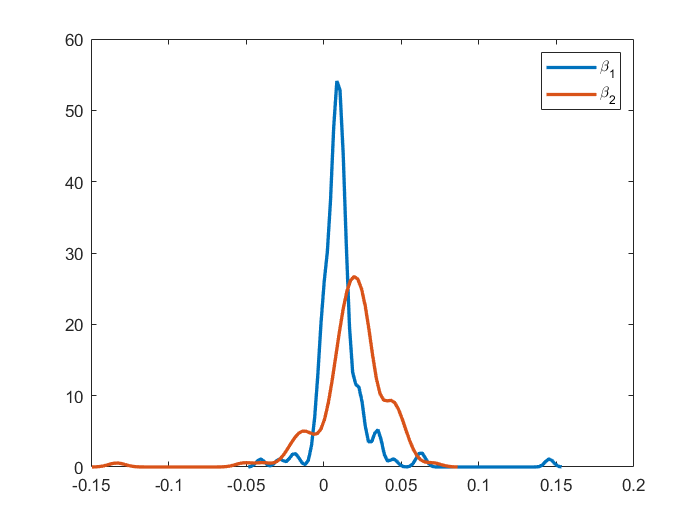}
         \caption{Density of the slopes in the broken-trend model}
         \label{fig_density_sb}
     \end{subfigure}
\caption{Density of individual estimates}
\label{fig_density_estimates}
\end{figure}

Assuming that individual trends are deterministic, rather than stochastic, we first model each individual grid through a linear-trend model of the form:
\begin{equation}
    T_{it} = \beta_{0i} + \beta_{1i} t + e_{it}, \hspace{0.3cm} i = 1,....,N, \hspace{0.3cm}, t = 1,....,T,
\label{eq_linear}
\end{equation}
where $e_{it} = \rho_{i} e_{it-1} + v_{it}$, $|\rho_{i}|<1$, and $v_{it}$ uncorrelated. For each sample period, we estimate Equation \ref{eq_linear} and generate density plots of $\hat{\beta_{1}}$. Panel (a) in Figure \ref{fig_density_estimates} shows that the density shifts to the right in more recent samples, reflecting the well-known acceleration in the global warming process \citep{Mann, ipcc, hansen}. Focusing solely on the estimated slopes for the sample period 1960-2022, the mean of the coefficient is higher for the set of grids that appear later in the record.

Acceleration in global warming, combined with the standard aggregation method A, can trivially generate the breaks in aggregated series even when the individual units follow a linear model. Consider the following extreme case. During the initial part of the record, a set of $N_{1}$ units with an average trend-slope $\bar{\beta_{1}}$ are observed. Assume that at a certain period, $TB$, a different set of $N_{2}$ units start to be observed with average slope $\bar{\beta_{2}}$, $\bar{\beta_{2}} > \bar{\beta_{1}}$. The slope of the average series is $\bar{\beta_{1}}$ before $TB$, and  $(N_{1}\bar{\beta_{1}} + N_{2}\bar{\beta_{2}})/(N_{1} + N_{2})$ after.\footnote{For the average computed using Method B the slope is $\bar{\beta_{1}}$ the full period.} Depending on the weights, the result of the ADF test can bias towards non-rejection. In the next section, we illustrate this point using simulations.  

Alternatively, we explore trend-breaks within individual grids in a model of the form:
\begin{equation}
    T_{it} = \alpha_{0i} + \alpha_{1i} D_{ui} + \gamma_{1i} t + \gamma_{2i}D_{ti} + e_{it}, \hspace{0.3cm} i = 1,....,N, \hspace{0.3cm}, t = 1,....,T,
    \label{eq_break}
\end{equation}
where $D_{ui} = 1\{t > TB_{i}\}$, $D_{ti} = 1\{t > TB_{i}\}\times(t-TB_{i})$, $TB_{i}$ is the period in which the structural break occurs for unit $i$, and $e_{it}$ follows the same structure as before. Out of the 155 grids observed continuously from 1880 to 2022, approximately 85\% contain a break in the trend around 1960. The densities of $\hat{\gamma_{1}}$ and $\hat{\gamma_{2}}$ plotted in Panel (b) of Figure \ref{fig_density_estimates} provide micro-level evidence of the warming acceleration phenomenon.

Aggregating individual grids with broken trends using either Method A or B results in averaged series with structural breaks, as those in Figure \ref{fig_average_1880_2022}. However, as discussed previously, due to the small number of grids involved in the computation of averages under method B, the signal of the break is not strong enough to wrongly drive the ADF test towards the conclusion of unit roots. For method A averages, the non-rejection of the ADF test can be attributed to both a higher signal, stemming from the utilization of more information each period, and the deepening of the break due to the acceleration in global warming that implies the inclusion of more grids in periods of higher trend. This hypothesis is supported by the simulations in the next section.

\section{Simulations}

The proposed 'micro-founded' explanations for the type of the trends in aggregate temperatures are validated heuristically using simulations. We consider two alternatives to simulate the non-missing indicator $I_{it}$:

\begin{itemize}
    \item Alternative 1: A fixed number of units are non-missing during the whole sample. A second group of series are missing during the first $T^{*}$ periods, and non-missing from $T^{*}$ on.
    \item Alternative 2: $I_{it}$ follows a Markov Switching (MS) process with two states (missing and non-missing) and transition-probability matrix $\mathbf{P}$.
\end{itemize}

First, we analyze the case where individual units follow a linear-trend model. Two groups of series are simulated:
\begin{equation}
    T_{it}^{1} = \beta_{0i}^{1} + \beta_{1i}^{1} t + e_{it}^{1}, \hspace{0.3cm} i = 1,....,N_{1}, \hspace{0.3cm} t = 1,....,T, 
\end{equation}
\begin{equation}
    T_{it}^{2} = \beta_{0i}^{2} + \beta_{1i}^{2} t + e_{it}^{2}, \hspace{0.3cm} i = 1,....,N_{2}, \hspace{0.3cm} t = 1,....,T, 
\end{equation}
where $\bar{\beta_{1}^{2}} > \bar{\beta_{1}^{1}}$ and $e_{it}^{j} = \rho_{i}^{j} e_{it-1}^{j} + v_{it}^{j}$, $j = 1,2$. Observation alternatives 1 and 2 assume that stations entering later in the average computation are chosen randomly from groups 1 or 2. To account for warming  and acceleration, we define observation alternatives 1* and 2* assuming that the series of group 2 (i.e. those with higher trend-slopes) appear as non-missing later in the record. 

The simulation parameters, including the transition probabilities, are set based on empirical evidence aiming to match the simulated and observed trends. Sample size is $T = 150$. The number of series are $N_{1} = 150$ and $N_{2} = 850$. The process is replicated $R = 1,000$ times. We impose that 10\% of units are observed from the initial period. Table \ref{tab_ur_sim_linear} presents the proportion of non-rejections of the ADF and KP tests for each observation alternative and aggregation method. Consistent with our hypothesis, even though the series are generated without unit root, the ADF test do not reject the null hypothesis in 100\% of the cases under observation alternatives 1* and 2* and aggregation method A. For aggregation Method B, the unit root is rejected in all cases. Moreover, the KP tests rejects the unit root in all cases.

\begin{table}[h]
\caption{Proportion of non-rejections in the unit root tests on simulated averages (linear-trend model)}
\centering
\begin{adjustbox}{width=0.6\textwidth}
\begin{threeparttable}
\begin{tabular}{lcccc}
\hline \hline
\multirow{2}{*}{\textbf{Alternative}} & \multicolumn{2}{c}{\textbf{Method A}} & \multicolumn{2}{c}{\textbf{Method B}} \\ & \textit{ADF-test} & \textit{KP-test} & \textit{ADF-test} & \textit{KP-test} \\ \hline 
1    & 0\%     & 0\%     & 0\%    & 0\%      \\
1*   & 100\%   & 0\%     & 0\%    & 0\%      \\
2    & 0\%     & 0\%     & 0\%    & 0\%      \\
2*   & 100\%   & 0\%     & 0\%    & 0\%      \\ \hline
\end{tabular}
\begin{tablenotes}
\item \tiny \textit{Notes:} The table reports the proportion of times that the null of unit root is non-rejected. Tests at 5\% of significance. Individual series simulated using the following calibration: $\beta_{0i}^{1} \sim \mathcal{N}(-0.7234, 0.4266^2)$, $\beta_{1i}^{1} \sim \mathcal{N}(0.0108, 0.0043^2)$,  $\beta_{0i}^{2} = -2.36$, $\beta_{1i}^{2} \sim \mathcal{N}(0.0271, 0.0145^2)$, $\rho_{i}^{1} \sim \mathcal{N}(0.264, 0.1176^2)$, $\rho_{i}^{2} \sim \mathcal{N}(0.155, 0.1747^2)$, and $v_{it}^{j} \sim \mathcal{N}(0, 3^2)$, $j = 1, 2$. These values are based on the estimates of fitting a linear model to individual grids over different samples.
\end{tablenotes} 
\end{threeparttable}
\end{adjustbox}
\label{tab_ur_sim_linear}
\end{table}

Next, lets consider the case where individual units contain one break in the trend function. Similar to the previous case, we simulate two groups of series of the form:
\begin{equation}
    T_{it}^{1} = \alpha_{0i}^{1} + \alpha_{1i}^{1} D_{ui}^{1} + \gamma_{1i}^{1} t + \gamma_{2i}^{1}D_{ti}^{1} + e_{it}^{1}, \hspace{0.3cm} i = 1,....,N_{1}, \hspace{0.3cm}, t = 1,....,T,
\end{equation}
\begin{equation}
    T_{it}^{2} = \alpha_{0i}^{2} + \alpha_{1i}^{2} D_{ui}^{1} + \gamma_{1i}^{2} t + \gamma_{2i}^{2}D_{ti}^{2} + e_{it}^{2}, \hspace{0.3cm} i = 1,....,N_{1}, \hspace{0.3cm}, t = 1,....,T,
\end{equation}
where the variables and parameters are defined in Equation \ref{eq_break}, $e_{it}^{j} = \rho_{i}^{j} e_{it-1}^{j} + v_{it}^{j}$, $j = 1,2$, and we impose $\bar{\gamma}_{2}^{2} > \bar{\gamma}_{2}^{1}$ to the capture warming acceleration. As in the previous case, alternatives 1* and 2* assume that the series in group 2 start to count for the average computation later.

The parameters are set based on empirical evidence, and the sample size, number of series, and replications are the same as before. Table \ref{tab_ur_sim_break} presents the proportion of non-rejections of the ADF and KP tests. Consistent with our hypothesis, when averages are computed using Method A, the unit root in ADF test is non-rejected in 100\% of the cases under all observation alternatives. For Method B, even though the original series contain a break and it is inherited by the aggregated series, the signal of the break is not strong enough to drive the test towards the non-rejection zone. However, if the proportion of units that are observed from the beginning of the record is increased, this signal eventually becomes stronger and the ADF test concludes the presence of unit roots.

\begin{table}[h]
\caption{Proportion of non-rejections in the unit root tests on simulated averages (model with broken-trend)}
\centering
\begin{adjustbox}{width=0.6\textwidth}
\begin{threeparttable}
\begin{tabular}{lcccc}
\hline \hline
\multirow{2}{*}{\textbf{Alternative}} & \multicolumn{2}{c}{\textbf{Method A}} & \multicolumn{2}{c}{\textbf{Method B}} \\ & \textit{ADF-test} & \textit{KP-test} & \textit{ADF-test} & \textit{KP-test} \\ \hline 
1    & 100\%     & 0\%     & 0\%    & 0\%      \\
1*   & 100\%   & 0\%     & 0\%    & 0\%      \\
2    & 100\%     & 0\%     & 0\%    & 0\%      \\
2*   & 100\%   & 0\%     & 0\%    & 0\%      \\ \hline
\end{tabular}
\begin{tablenotes}
\item \tiny \textit{Notes:} The table reports the proportion of times that the null of unit root is non-rejected. Tests at 5\% of significance. Individual series simulated using the following calibration: $\alpha_{0i}^{1}, \alpha_{0i}^{2}  \sim \mathcal{N}(-0.6061, 0.5642^2)$,  $\alpha_{1i}^{1}, \alpha_{1i}^{2}  \sim \mathcal{N}(-0.5524, 0.6899^2)$, $\gamma_{1i}^{1}, \gamma_{1i}^{2}  \sim \mathcal{N}(0.0110, 0.0174^2)$, $\gamma_{2i}^{1}, \sim \mathcal{N}(0.0182, 0.0233^2)$, $\gamma_{2i}^{2}, \sim \mathcal{N}(0.0271, 0.0145^2)$, $\rho_{i}^{j} \sim \mathcal{N}(0.1101, 0.0899^2)$, and $v_{it}^{j} \sim \mathcal{N}(0, 3^2)$, $j = 1, 2$. These values are based on the estimates of fitting a model allowing for a break in the trend to individual grids.
\end{tablenotes} 
\end{threeparttable}
\end{adjustbox}
\label{tab_ur_sim_break}
\end{table}

\section{Conclusions}

Aggregate and micro-founded evidence do not support the hypothesis of stochastic trends in temperature. In fact, our evidence suggests that temperatures averages are stationary around a non-linear trend, with the non-linearity being modelled as a one-time break in the trend function. The break is attributed to a combination of individual grid breaks and the standard aggregation method under warming acceleration. Moreover, the aggregation method is relevant to bias ADF-tests towards the non-rejection zone by amplifying the signal of the break. Our findings carry important empirical implications for studies on the detection, attribution, impact, and forecasting of global warming. Radiative forcing from anthropogenic greenhouse gases (such as CO2, methane, and nitrous oxide) is modelled as an I(1), or even I(2) process. It introduces a problem of 'unbalanced' regressions if the aim is to establish human influence on global warming using cointegration methods. A similar issue is present on impact studies relating economic growth and temperature data. A valid alternative includes exploring co-trending approaches (see \cite{Chenetal2022}) or adopting statistical methods robust to the types of trends.

\clearpage
\begin{spacing}{0.8}
\bibliographystyle{chicago}
\bibliography{References}
\end{spacing}

\end{document}